\documentclass[lettersize,journal]{IEEEtran}
\usepackage{amsmath}
\usepackage{amssymb}
\usepackage{stfloats}
\usepackage{url}
\usepackage[square,sort&compress,comma,numbers]{natbib}
\usepackage{cleveref}
\usepackage{array}
\usepackage{adjustbox}
\usepackage{dirtytalk}
\usepackage{soul}
\usepackage{placeins} 
\usepackage{afterpage}
\usepackage{bm}
\usepackage{upgreek}
\usepackage{mathtools}

\usepackage[symbol]{footmisc}
\renewcommand{\thefootnote}{\fnsymbol{footnote}}

\graphicspath{{figures/}} 

\begin{document}

\title{Centralised rehearsal of decentralised cooperation: Multi-agent reinforcement learning for the scalable coordination of residential energy flexibility}

\author{
\IEEEauthorblockN{Flora Charbonnier\textsuperscript{\footnotemark*}}
\IEEEauthorblockN{Bei Peng\textsuperscript{\footnotemark†}}
\IEEEauthorblockN{Thomas Morstyn\textsuperscript{\footnotemark*}} \and
\IEEEauthorblockN{Malcolm McCulloch\textsuperscript{\footnotemark*}}
}

\markboth{Journal of \LaTeX\ Class Files,~Vol.~14, No.~8, August~2021}%
{Shell \MakeLowercase{\textit{et al.}}: A Sample Article Using IEEEtran.cls for IEEE Journals}


\maketitle

\footnotetext[1]{Department of Engineering Science, University of Oxford}
\footnotetext[2]{Department of Computer Science, University of Liverpool}

\renewcommand*{\thefootnote}{\arabic{footnote}}

\begin{abstract}
This paper investigates how deep multi-agent reinforcement learning can enable the scalable and privacy-preserving coordination of residential energy flexibility. The coordination of distributed resources such as electric vehicles and heating will be critical to the successful integration of large shares of renewable energy in our electricity grid and, thus, to help mitigate climate change. The pre-learning of individual reinforcement learning policies can enable distributed control with no sharing of personal data required during execution. However, previous approaches for multi-agent reinforcement learning-based distributed energy resources coordination impose an ever greater training computational burden as the size of the system increases. We therefore adopt a deep multi-agent actor-critic method which uses a \emph{centralised but factored critic} to rehearse coordination ahead of execution. Results show that coordination is achieved at scale, with minimal information and communication infrastructure requirements, no interference with daily activities, and privacy protection. Significant savings are obtained for energy users, the distribution network and greenhouse gas emissions. Moreover, training times are nearly 40 times shorter than with a previous state-of-the-art reinforcement learning approach without the factored critic for 30 homes.
\end{abstract}


\section{Introduction}
\IEEEPARstart{T}{his} paper tackles the problem of computational scalability in coordinating residential energy flexibility while maintaining privacy. To achieve this, we propose the use of cooperative multi-agent reinforcement learning (MARL) with a centralised but factored critic.

The decarbonisation of power supply, concurrent with the electrification of heat and transport, is urgently required to mitigate the climate crisis. To this end, large shares of non-dispatchable renewable energy must supply the power grid, just as the loads to be managed increase, thus heightening the need for demand-side flexibility coordination. Residential energy has substantial such flexibility potential, particularly as private transport and heat provision are electrified. For example, in the UK, the residential sector already represents about one third of overall electricity demand and up to 60\% of peak demand, when the value of flexibility is highest \citep{Torriti2022}. 

The cost of two-way information and distributed computation infrastructure is a hurdle in the way of this coordination, as the large overall residential potential is split into millions of homes with limited individual value \citep{Leautier2019}. Moreover, users may distrust interference with their daily activities and privacy \citep{Pumphrey2020,Bugden2019}. The question of computational scalability of real-time control of millions of homes is also unresolved \citep{Charbonnier2022_Review}. 

The use of reinforcement learning (RL) has therefore been proposed to coordinate distributed energy resources (DERs). The decentralised execution of pre-learned policies can overcome these challenges whilst improving robustness and removing the need for a convex representation of complex systems \citep{Charbonnier2022_Review}. However, most existing works have relied on the centralised \citep{ONeill2010, Antonopoulos2020, Darby2020, Chen2019, Lu2019, Kim2016, Babar2018, Vaya2014, Ye2020, Dauer2013, Sun2015, Kim2020, Claessens2013, Zhang2017, Dusparic2015, Hurtado2018} or bilateral \citep{Taylor2014} sharing of personal information. This may raise issues around communication infrastructure costs, security, privacy, computational scalability, and vulnerability to biased information \citep{Charbonnier2022_Review, Guerrero2020,Morstyn2020_P2P}. 

In implicit coordination strategies, on the other hand, DERs are coordinated without sharing personal data \citep{Charbonnier2022_Review}. Advantages include reduced complexity and costs of the ICT infrastructure, enhanced privacy, self-control and acceptability for users, robustness against failures, and reliability \citep{Guerrero2020,Mai2021}. In RL-based implementations, the data access and computational burden needs are shifted to a pre-training phase, when agents learn individual policies in a shared environment. A fully decentralised execution at the home level then only depends on local information. This multi-agent task can be modelled as a \textit{decentralised partially observable Markov decision process} (Dec-POMDP) \citep{oliehoek2016concise}. Yet, a competitive approach \cite{Cao2019, Yang2019}, where each agent only seeks to maximise their own utility, may lead to sub-optimal outcomes, negatively impacting the grid and individuals. The peak may be merely displaced, with overloads on upstream transformers \cite{Crozier2018_Mitigating}. The addition of a network management term to the individual competitive rewards functions was therefore suggested \citep{Pigott2022}. In \citep{Charbonnier2022_MARL}, full cooperation is achieved as independent learners concurrently learn to maximise a shared multi-objective function. 

The computational scalability of the training phase for cooperative DER coordination remains a challenge. To improve learning efficiency, we seek to adopt a MARL algorithm that fully exploits the benefits of the \textit{centralised training with decentralised execution} (CTDE) \citep{oliehoek2008optimal} paradigm. In CTDE, due to partial observability or  communication constraints, each agent must learn a \textit{decentralised} policy conditioned only on local observations. However, the training itself can be \emph{centralised} in a simulated environment, with access to additional information about the environment (e.g., global state) and other agents. \Cref{tab:CTDE} summarises three common classes of cooperative MARL approaches used to solve a Dec-POMDP. 

The simplest type is \emph{independent learning}, where each agent treats other agents as part of the environment and learns independently. For instance, in independent Q-learning (IQL) \citep{Tan1993}, each agent $i$ learns a \textit{decentralised} action-value function $Q_i$ based only on individual observations and actions. This method is limited as it cannot explicitly represent interactions between agents. It is also prone to instability as each agent’s learning is confounded by the learning and exploration of others and the non-stationarity of the environment \citep{Rashid2018, Matignon2012}. 



In contrast, one category of CTDE algorithms is the \textit{centralised multi-agent policy gradient} method \citep{Gupta2017, Foerster2018, Lowe2017, Lowe2020}, in which each agent learns a centralised critic with a decentralised actor. The \textit{centralised and monolithic} critic utilises the global state and the actions of all agents, which are only available during centralised training, to estimate the \textit{centralised} action-value function $Q_{tot}$. Compared to $Q_i$, $Q_{tot}$ improves coordination by capturing the interdependent effects of all agents' actions and guiding the optimisation of decentralised policies. Yet, learning a satisfactory estimate of $Q_{tot}$ can be impractical since it directly conditions on the global state and joint action, which can grow exponentially with the number of agents.   

\textit{Value function factorisation} methods \citep{Rashid2018, rashid2020weighted, wang2021qplex, Hu2020, Qiu2021} constitute another category of CTDE algorithms. The centralised action-value function $Q_{tot}$ is represented as a mixing function of individual action-value functions $Q_i$ to enable easy decentralisation and enhance scalability. Value decomposition has been shown to be an effective approach in most environments and shares the major advantages of centralised training, especially in environments with dense rewards \citep{Papoudakis2020}, such as in DER coordination. However, how to best represent and learn $Q_{tot}$ is still an open question. 

 \begin{table}[h!]
\begin{center}
\begin{tabular}{m{3cm} c }
\hline
MARL Approach & Schematic illustration \\ 
\hline
 & \\
Independent learning \citep{Tan1993, Mnih2013, Omidshafiei2017,Foerster2017, Charbonnier2022_MARL} &
\adjustbox{valign=c}{\includegraphics[width=0.13\textwidth]{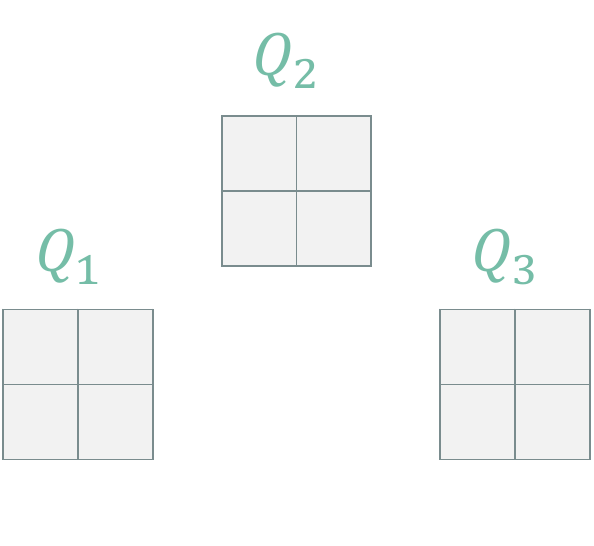}} \\ 

Centralised multi-agent policy gradient \citep{Foerster2018, Lowe2017, Lowe2020, Gupta2017} & 
\adjustbox{valign=c}{\includegraphics[width=0.11\textwidth]{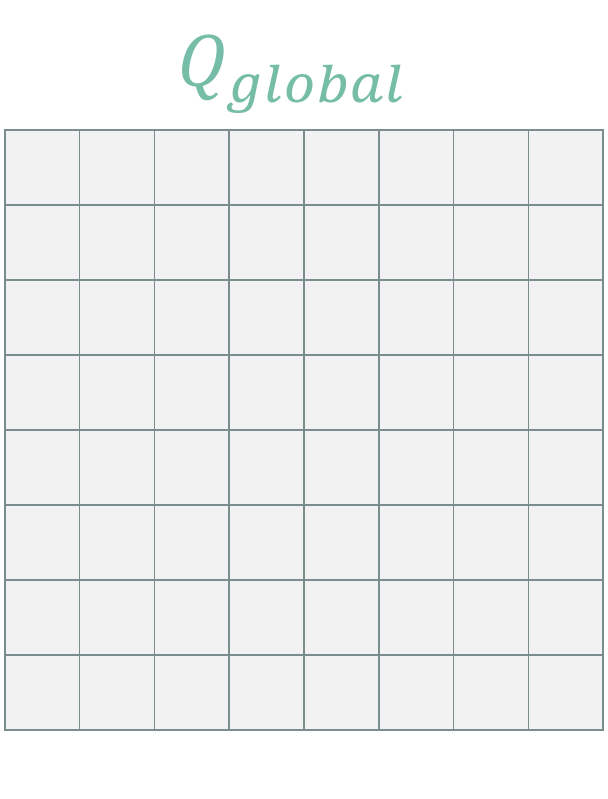}} \\ 

Value function factorisation \citep{Rashid2018, rashid2020weighted, wang2021qplex, Hu2020, Qiu2021, Peng2021} &
\adjustbox{valign=c}{\includegraphics[width=0.15\textwidth]{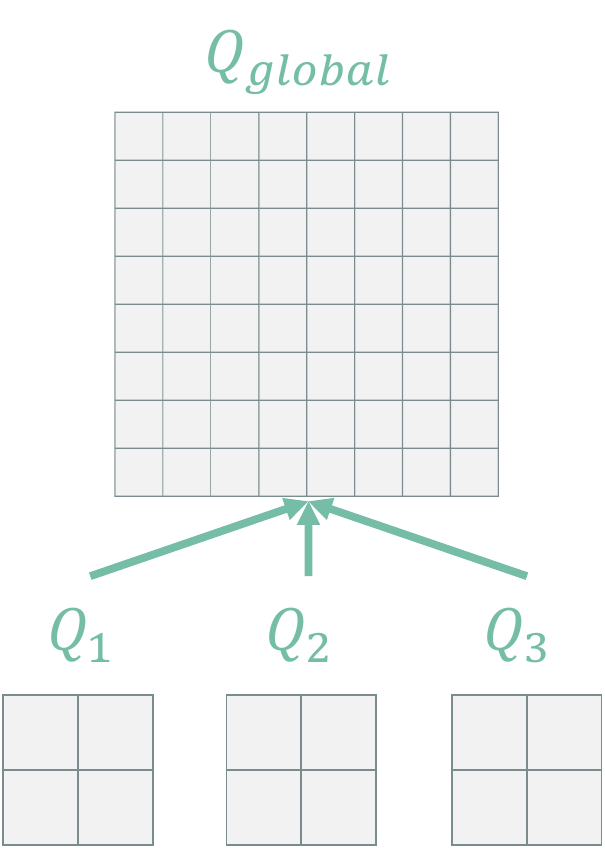}} \\ 

&\\
\hline

\end{tabular}
\caption[Three Common Classes of Multi-agent reinforcement learning (MARL) approaches to solve a decentralised partially observable Markov decision process (Dec-POMDP)]{Three Common Classes of Multi-agent reinforcement learning (MARL) approaches to solve a decentralised partially observable Markov decision process (Dec-POMDP)\protect\footnotemark[2].}

\label{tab:CTDE}
\end{center}
\end{table}

\setcounter{footnote}{0}

Here, to solve the residential energy flexibility coordination problem, we propose the use of the factored multi-agent centralised policy gradient (FACMAC) approach \citep{Peng2021}. FACMAC is a multi-agent actor-critic method that learns a single \textit{centralised but factored} critic\footnote{For the remainder of this paper, we will use \say{FACMAC} and \say{centralised but factored critic approach} interchangeably.}, which factors the centralised action-value function $Q_{tot}$ as a non-linear monotonic function of individual action-value functions $Q_i$. It therefore combines the advantages of both centralised multi-agent policy gradient and value function factorisation methods.
Whilst this could allow home energy management systems to learn more coordinated behaviour when privately operating their electric vehicles (EVs), heating, and other flexible loads, this potential has thus far not been investigated.

The contribution of this paper is thus to build a bridge between the fields of DER coordination and MARL to reduce the computational burden associated with residential energy flexibility coordination. Using a multi-agent actor-critic framework that learns decentralised policies with a centralised but factored critic addresses the pitfalls of both the poor coordination performance of independent learners and the intractability at scale of optimisations or centralised critic estimation. To enhance the performance of DER coordination, we further improve the centralised but factored critic methodology by incorporating two techniques: convolutional neural networks \citep{Lecun1998} and hysteretic learning \citep{Matignon2007}. Furthermore, we extend the supervised loss approach from single-agent RL \citep{Hester2018} to the MARL case, to combine computationally \say{expensive} training data obtained from system optimisations results and \say{cheap} environment exploration to guide the learning of multi-agent cooperation. 

\footnotetext[2]{Note that the action-value functions are represented as tables (as would be the case in a tabular Q-learning implementation) for illustration purposes, though these can be estimated using function approximators such as neural networks.}

\setcounter{footnote}{2}

This work freely provides a local energy environment and MARL testing and benchmarking framework on GitHub\footnote{\url{https://github.com/floracharbo/MARL_local_electricity}}. MARL research is generally conducted on virtual benchmark game environments designed explicitly for evaluating and developing MARL algorithms. If MARL is to leave the laboratory setting, benchmark environments must tackle real-world problems such as DER coordination.

The rest of this paper is organised as follows. In \Cref{sec:system}, the testing environment is presented that includes household-level modelling of EVs, space heating, flexible loads and photovoltaic (PV) generation. We then describe the MARL methodologies used to coordinate home energy consumption in this system in \Cref{sec:methodology}. Their performance is assessed in experiments conducted using real-life data in \Cref{sec:results}. Finally, we conclude in \Cref{sec:conclusion}.

\section{Problem description}\label{sec:system}

In this section, the variables, objective function and constraints of
the problem are described. This sets the frame for the application of the MARL algorithms presented in \Cref{sec:methodology}. This model is modular, and as such energy resources in each home could readily be removed, modified (e.g., using other car models) and added (e.g., modelling smart hot water tanks) without reducing its compatibility with the MARL methods. This work uses the model presented in \citep{Charbonnier2022_MARL}.

\subsection{Local energy system model variables}
\begin{figure*}[!t]
\begin{center}
\includegraphics[width=0.9\linewidth]{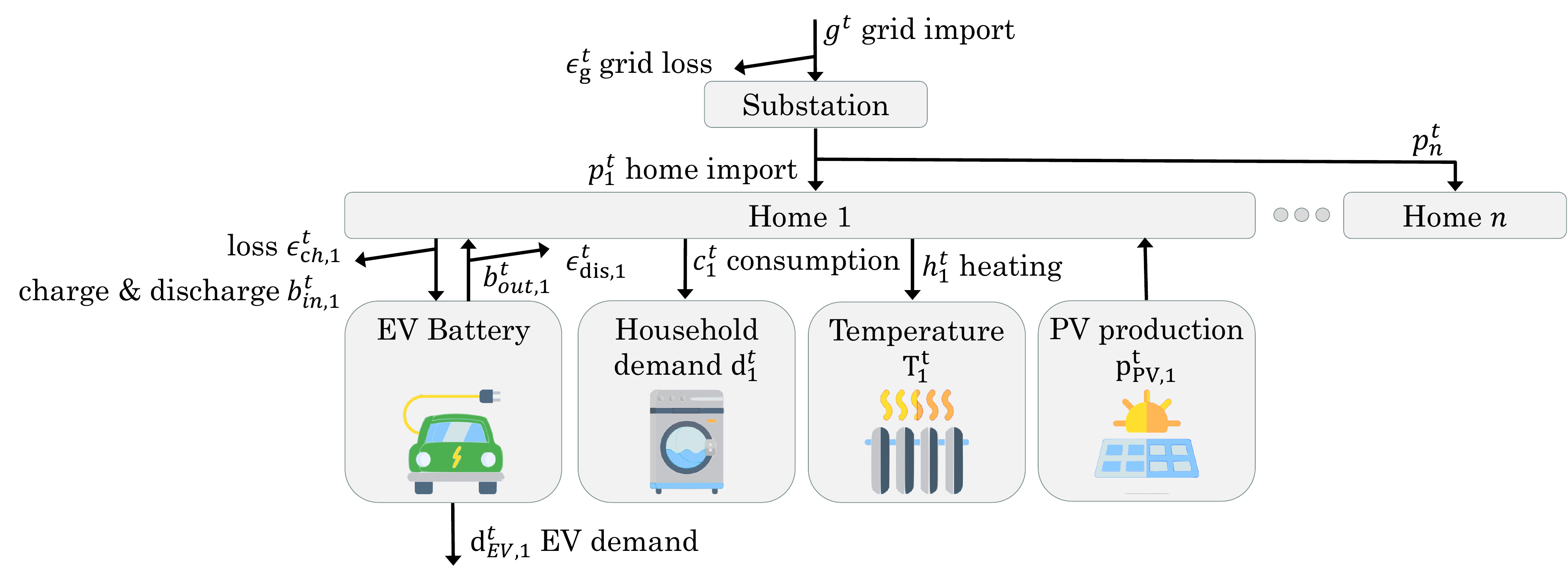}
\caption{Local energy system model. Homes have vehicle-to-grid (V2G)-enabled electric vehicles, flexible household demand, electric heating, and PV generation. Energy balances apply at the asset-, home- and substation levels.}
\label{fig:energy_balance}
\end{center}
\end{figure*}

The local system model is illustrated in \Cref{fig:energy_balance}. Participants under a secondary substation have a V2G-enabled EV, a PV panel, electric space heating and generic flexible loads. We consider a set of time steps $t \in \mathcal{T} = \{t_0,...,t_\textrm{end}\}$ and a set of homes $i \in \mathcal{P} = \{1,...,n\}$. Decision variables are \emph{italicised} and input data are written in roman. Energy units are used unless specified otherwise.

The EV at-home availability $\upmu_i^t$ (1 if available, 0 otherwise), EV demand for required trips $\textrm{d}_{\textrm{EV},i}^t$, household electric demand $\textrm{d}_i^{t}$,  PV production $\textrm{p}_{\textrm{PV},i}^t$, external temperature $\textrm{T}_{\textrm{e}}^t$ and solar heat flow rate $\upphi^t$ are specified as inputs for $t \in \mathcal{T}$ and $i \in \mathcal{P}$.

The local decisions are the battery charge $b_{\textrm{in},i}^t$ and discharge $b_{\textrm{out},i}^t$, the electric heating consumption $h_i^t$ and the household consumption $c_i^t$. These have local and system impacts. Local impacts include battery energy levels $E_i^t$,  losses $\epsilon_{\textrm{ch},i}^t$ and $\epsilon_{\textrm{dis},i}^t$, home imports $p_i^t$, building mass temperatures $T_{\textrm{m},i}^t$ and indoor air temperatures $T_{\textrm{air},i}^t$. System impacts arise through the costs of total grid import $g^t$ and distribution network use. Distribution network losses and reactive power flows are not included in this work.

\subsection{Objective function}\label{sec:objfunc}
All homes cooperate to minimise the sum of grid ($c_\textrm{g}^t$), distribution ($c_\textrm{d}^t$) and storage ($c_\textrm{s}^t$) costs.

\begin{equation}
\max F = \sum_{\forall t \in \mathcal{T}}{\hat{F}_t} = \sum_{\forall t \in \mathcal{T}}{- (c_\textrm{g}^t + c_\textrm{d}^t + c_\textrm{s}^t )}
\end{equation}

\begin{equation}
c_\textrm{g}^t = \textrm{C}_\textrm{g}^t \left( g^t + \epsilon_g \right)
\end{equation}

The grid cost coefficient $\textrm{C}_\textrm{g}^t$ is the sum of the grid electricity price and the product of the carbon intensity at time $t$ and the social cost of carbon, which reflects the long-term societal cost of emitting greenhouse gases \citep{ParryM}. The impacts of local decisions on upstream energy prices are neglected.

\begin{equation}
c_\textrm{d}^t = \textrm{C}_\textrm{d}\sum_{i \in \mathcal{P}}{\max\left(- p_i^t,0\right)}
\end{equation}
\noindent Distribution costs $c_\textrm{d}^t$ are proportional to the distribution charge $\textrm{C}_\textrm{d}$ on exports. The resulting price spread between individual imports and exports decreases network constraint violation risks by incentivising the use of local flexibility first \citep{Morstyn2020_IntegratingP2P}.
\begin{equation}
c_\textrm{s}^t = \textrm{C}_\textrm{s}\sum_{i \in \mathcal{P}}{\left(b_{\textrm{in},i}^t + b_{\textrm{out},i}^t\right)}
\end{equation}

\noindent Storage battery depreciation costs $c_\textrm{s}^t$ assume a uniform energy throughput degradation rate $\textrm{C}_\textrm{s}$ \citep{DufoLopez2014}.

\subsection{Constraints}
The constraints for steps $\forall \ t \in \mathcal{T}$ and homes $\forall \ i \in \mathcal{P}$ are:

\begin{itemize}
\item Substation-, home- and EV battery-level energy balances -- for $\upeta_\textrm{ch}$ and $\upeta_\textrm{dis}$ the charge and discharge efficiencies:
\begin{equation}
\sum_{i \in \mathcal{P}}{p_i^t} = g^t
\end{equation}
\begin{equation}
	p_i^t = c_i^t + h_i^t + \frac{b_{\textrm{in},i}^t}{\upeta_\textrm{ch}} - {\upeta_\textrm{dis}} b_{\textrm{out},i}^t - \textrm{p}_{\textrm{PV},i}^t 
\end{equation}

\begin{equation}\label{eq:bat_balance}
	E_i^{t+1} = E_i^t  + b_{\textrm{in},i}^t - b_{\textrm{out},i}^t - \textrm{d}_{\textrm{EV},i}^t 
\end{equation}

\item Battery charge and discharge constraints -- Let $\textrm{E}_0$,  $\underline{\textrm{E}}$ and $\overline{\textrm{E}}$ be the initial, minimum and maximum battery energy levels, and $\overline{\textrm{b}_\textrm{in}}$ and $\overline{\textrm{b}_\textrm{out}}$ the maximum charge and discharge per time step:
\begin{equation}\label{eq:bat_0}
	 \textrm{E}_0 = E_i^{t_0}  = E_i^{t_\textrm{end}} + b_{\textrm{in},i}^{t_{\textrm{end}}} - b_{\textrm{out},i}^{t_{\textrm{end}}} - \textrm{d}_{\textrm{EV},i}^{t_{\textrm{end}}} 
\end{equation}
\begin{equation}\label{eq:bat_bounds}
	\upmu_i^t\underline{\textrm{E}}_i \leq E_i^t \leq \overline{\textrm{E}}_i
\end{equation}
\begin{equation}\label{eq:bat_in}
	b_{\textrm{in},i}^t \leq \upmu_i^t \overline{\textrm{b}_\textrm{in}}
\end{equation}
\begin{equation}\label{eq_bat_out}
	b_{\textrm{out},i}^t \leq \overline{\textrm{b}_\textrm{out}}
\end{equation}

\item Consumption flexibility -- Demand $\textrm{d}_{i,k}^{t_\textrm{D}}$ is met by the sum of partial consumptions $\hat{c}_{i,k,t_\textrm{C},t_\textrm{D}}$ at time $t_\textrm{C}$ by prosumer $i$ for load of type $k$ (fixed or flexible) demanded at $t_\textrm{D}$. The flexibility boolean $\textrm{f}_{i,k,t_\textrm{C},t_\textrm{D}}$ indicates if time $t_\textrm{C}$ lies within the acceptable range $\{t_\textrm{D},...,t_{\textrm{D}+\textrm{n}_\textrm{flex}}\}$ to meet $\textrm{d}_{i,k}^{t_\textrm{D}}$.
\begin{equation}\label{eq:demandmet}
	\sum_{t_\textrm{C}\in\mathcal{T}}{\hat{c}_{i,k,t_\textrm{C},t_\textrm{D}} \textrm{f}_{i,k,t_\textrm{C},t_\textrm{D}}} = \textrm{d}_{i,k}^{t_\textrm{D}} 
\end{equation}
\item Consumption -- the total consumption at time $t_\textrm{C}$ is the sum of all partial consumptions $\hat{c}_{i,k,t_\textrm{C},t_\textrm{D}}$:
\begin{equation}\label{eq:totalcons}
	\sum_{t_\textrm{D}\in\mathcal{N}}{\hat{c}_{i,k,t_\textrm{C},t_\textrm{D}}}= c_{i,k}^{t_\textrm{C}} 
\end{equation}

\item Heating -- A Crank-Nicholson scheme \citep{ISO2007} was reformulated as a linear recursive expression, with $\upkappa$ a 2x5 matrix of temperature coefficients, and $\underline{\textrm{T}}_i^t$ and $\overline{\textrm{T}}_i^t$ lower- and upper-temperature bounds:
    \begin{equation}\label{eq:main_heating}
\begin{bmatrix}
T_{\textrm{m},i}^{t+1}\\
T_{\textrm{air},i}^{t+1} 
\end{bmatrix}
 = \upkappa 
 \begin{bmatrix}
1,
T_{\textrm{m},i}^{t},
\textrm{T}_{\textrm{e}}^t,
\upphi^t,
h_i^t
\end{bmatrix}^\intercal
\end{equation}

\begin{equation}\label{eq:temp_limits}
\underline{\textrm{T}}_i^t \leq T_{\textrm{air},i}^t \leq \overline{\textrm{T}}_i^t
\end{equation}

\item Non-negativity constraints: 
\begin{equation}
	c_i^t, h_i^t,E_i^t, b_{\textrm{in},i}^t, b_{\textrm{out},i}^t, \hat{c}_{i,l,t_\textrm{C},t_\textrm{D}} \geq 0
\end{equation}

\end{itemize}

\section{Reinforcement learning strategy formulation}\label{sec:methodology}
We now present the MARL methodologies which allow independent agents (i.e., homes) to learn to make individual decisions to maximise the total discounted reward, which is equivalent to maximising the statistical expectation of the objective function for the energy system described in \Cref{sec:objfunc}.

The MARL policies are trained in offline simulations prior to potential online executions in the electricity grid, so agents do not trial unsuccessful actions with real-life impacts during learning. Moreover, this means that the computation burden is taken prior to implementation, and homes would merely need to apply pre-learned policies, avoiding the computational challenges of large-scale real-time control in power system applications.

\subsection{Reinforcement learning}
RL is a machine learning paradigm where an agent aims to solve a sequential decision-making problem by directly interacting with an uncertain environment \citep{Sutton1998}. The goal of the RL agent is to learn a policy (i.e., a sequence of actions) that maximises its long-term expected total reward. 

We consider a fully cooperative multi-agent task in which a team of agents interacts with the same environment to achieve some common goal, which can be modelled as a Dec-POMDP \citep{oliehoek2016concise}. The Dec-POMDP consists of a tuple $G = \langle \mathcal{N} , S, A, P, r, O, \gamma \rangle$. Here $\mathcal{N} \equiv \{1, ..., n\}$ denotes the finite set of agents and $s \in S$ describes the true state of the environment. At each time step, each agent $i \in \mathcal{N}$ selects a discrete or continuous action $a_i \in A$, forming a joint action $\bm{a} \in \bm{A} \equiv A^n$. The environment then produces a transition to the next state $s^{\prime}$ according to the state transition function $P(s^{\prime}|s, a) : S \times A \times S \to [0, 1]$ and a team reward $r(s, a)$. $\gamma \in [0, 1)$ is a discount factor. Due to the partial observability, each agent $i \in \mathcal{N}$ draws an individual partial observation $o_i \in \Omega$ from the observation kernel $O(s, i)$. Each agent learns a policy $\mu_i(\tau_i)$, conditioned only on its local action-observation history $\tau_i \in T \equiv (\Omega \times A)^*$, which may be stochastic or deterministic. The joint policy $\mu$ induces a joint action-value function: $Q^\mu(s_t, a_t) = \mathbb{E}_{s_t+1:\infty,a_t+1:\infty}[R_t|s_t, a_t]$, where $R_t = \sum_{i=0}^\infty{\gamma^ir_{t+i}}$ is the discounted return.

RL methods were initially developed for single agents, rather than for large numbers of agents under partial observability in stochastic environments. Further methodological innovation was therefore needed for multiple such agents to learn to cooperate \citep{Charbonnier2022_MARL}.

\subsection{Optimisation-informed independent learning}
Some of these challenges have been mitigated in \emph{optimisation-informed independent learning} \citep{Charbonnier2022_MARL} by generating training data using \say{omniscient} convex optimisations. Agents thus concurrently learn by interacting with a shared environment using individual, decentralised fixed-size Q-tables, guided by the action choice of the agents in the convex optimisation solutions. These stable, consistent solutions align with the global optima Pareto and can successfully act as a coordination mechanism \citep{Charbonnier2022_MARL}. 

Additional baselining simulations, which compare total rewards to that if each agent took a baseline, inflexible action, can further improve \emph{learnability}\footnote{\say{the sensitivity of an agent’s utility to its own actions as opposed to actions of others, which is often low in fully cooperative Markov games} \citep{Matignon2012}}. Each agent thus learns from their marginal reward, i.e., the difference in total instant rewards $r^t$ between that if agent $i$ selects the greedy action and that if it selects the default action is used to update $Q^\textrm{diff}$. The default reward $r^t_{a_{i}=a_\textrm{default}}$, where all agents perform their greedy action apart from agent $i$, which performs the default action, is obtained by an additional simulation.
 \begin{equation}
Q(s_i^t, a_i^t)\leftarrow Q(s_i^t,a_i^t) + \alpha \delta
\end{equation}
where 
\begin{equation}
\delta = \left(r^t - r^t_{a_{i}=a_\textrm{default}}\right) + \gamma V^\textrm{diff}(s_i^{t+1}) - Q^\textrm{diff}(s_i^t, a_i^t)
\end{equation}

However, although this allows for stable coordination performance as the number of agents increases, these coordination mechanisms come at a computational cost at scale. 

\subsection{Centralised but factored critic}
\begin{figure*}[!t]
\begin{center}
\includegraphics[width=0.8\linewidth]{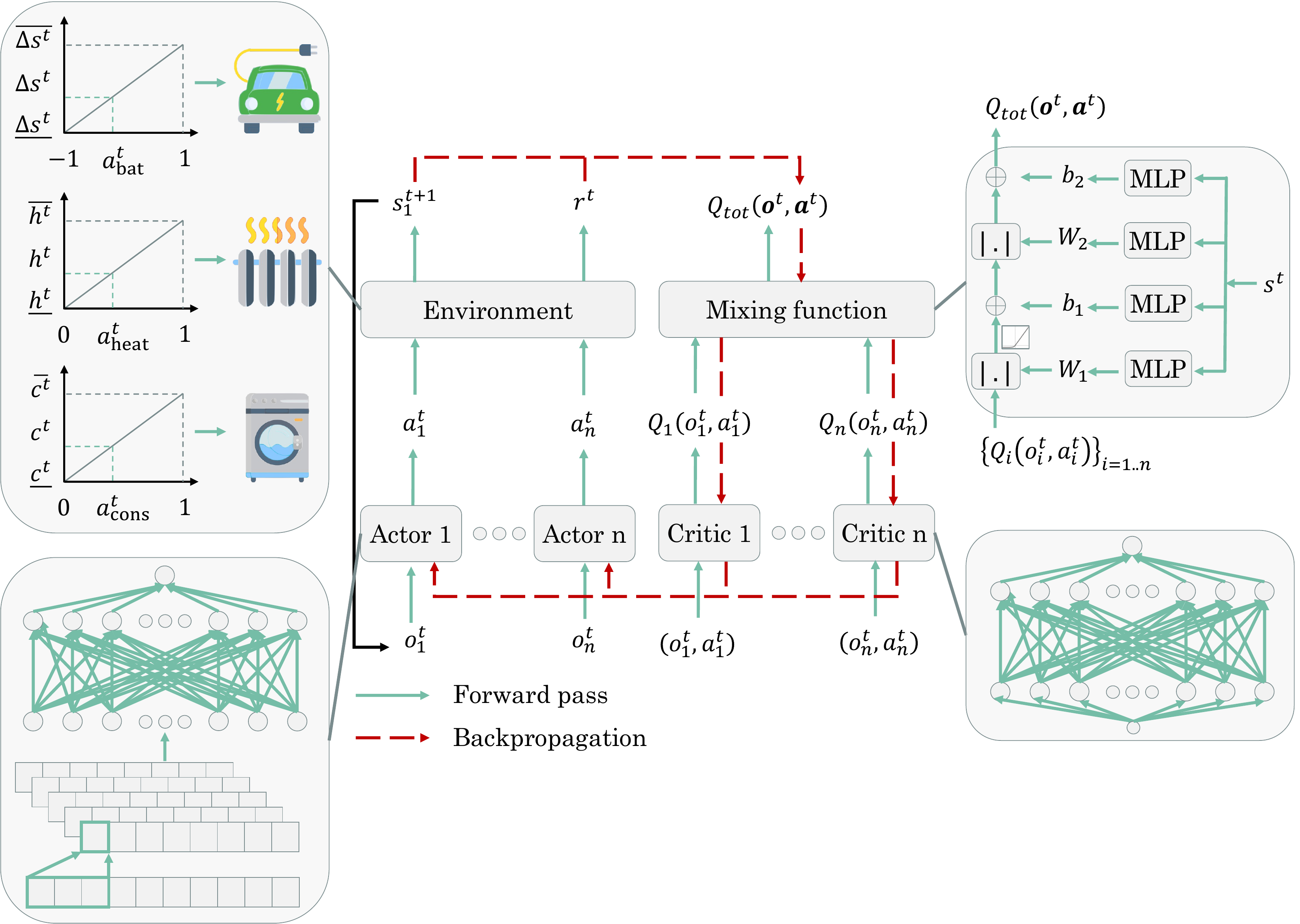}

\end{center}
\caption{The overall FACMAC architecture. 
For each agent $i$, there is one actor network that selects individual action $a_i^t$ based on local observation $o_i^t$. This action is inputted into the local energy environment to obtain the next state $s_i^{t+1}$ and reward $r^t$.
For each agent $i$, there is one critic network that estimates the individual action-value function $Q_i$, which is then combined into the centralised action-value function $Q_{tot}$ via a non-linear monotonic mixing function approximated by a mixing network with non-negative weights.
The actor network includes a convolutional layer followed by two hidden layers. The critic network is a linear network with one hidden layer. 
}
\label{fig:facmac}
\end{figure*}

\Cref{fig:facmac} schematically illustrates the structure of the modified factored multi-agent centralised policy gradients (FACMAC) algorithm \citep{Peng2021}. Like multi-agent deep deterministic gradients (MADDPG) \citep{Lowe2017}, a popular multi-agent actor-critic method, FACMAC uses deep deterministic policy gradients to learn policies. However, FACMAC learns a centralised but factored critic, which combines per-agent utilities into the centralised action-value function via a non-linear monotonic function, as in QMIX \citep{Rashid2018}. This enables more scalable learning of the centralised critic. 
Moreover, FACMAC uses a new centralised policy gradient estimator that optimises over the entire joint action space to allow for more coordinated policy changes and fully reap the benefits of a centralised critic.

In FACMAC, all agents share a centralised critic $Q^\mu_\textrm{tot}$ that is factored as:
\begin{equation}
    Q^\mu_\textrm{tot}(\bm{\tau}, \bm{a}, s; \bm{\phi}, \psi) = g_\psi(s, \{Q_i^{\mu_i}(\tau_i, a_i; \psi_i)\}^n_{i=1})
\end{equation}

where $\bm{\phi}$ and $\psi$ are parameters of the centralised action-value function $Q^\mu_\textrm{tot}$ and agent-wise utilities $Q_i^{\mu_i}$, respectively, and $g_\psi$ is a non-linear
monotonic function parameterised as a mixing network with parameters $\psi$, as in QMIX \citep{Rashid2018}. To evaluate the policy, the centralised but factored critic is trained by minimising the following loss:
\begin{equation}
    \mathcal{L}(\bm{\phi}, \psi) = \mathbb{E}_\mathcal{D}[(y^\textrm{tot} - Q^\mu_\textrm{tot}(\bm{\tau}, \bm{a}, s; \bm{\phi}, \psi)^2]
\end{equation}
where $y^\textrm{tot}=r + \gamma Q^\mu_\textrm{tot}(\bm{\tau'}, \bm{\mu(\tau'; \theta^-)}, s'; \bm{\phi^-}, \psi^-)$, $\mathcal{D}$ is the replay buffer, and $\bm{\theta^-}$, $\bm{\phi^-}$ and $\psi^-$ are the parameters of the target actors, critic, and mixing network, respectively. 

To update the decentralised policy of each agent, a centralised policy gradient estimator is used to optimise over the entire joint action space:
\begin{equation}
    \nabla_\theta J(\bm{\mu}) = \mathbb{E}_\mathcal{D}[\nabla_\theta\bm{\mu}\nabla_{\bm{\mu}}Q_\textrm{tot}^{\bm{\mu}}(\bm{\tau}, \mu_1(\tau_1), ..., \mu_n(\tau_n), s)]
\end{equation}

where $\bm{\mu}=\{\mu_1(\tau_1; \theta_1), ..., \mu_n(\tau_n; \theta_n)\}$ is the set of all agents' current policies, and all agents share the same actor network parameterised by $\theta$.

We build on this work by making three additional methodological contributions:

\subsubsection{Convolutional networks} A convolutional layer with kernel size 3 allows for improved feature extraction on both the actor networks when learning action time series selection for the day ahead.

\subsubsection{Hysteretic learning} Hysteretic learners are optimistic learners that use a higher learning rate for increasing Q-values than for decreasing Q-values, which helps to reduce oscillations in the learned policy due to actions chosen by other agents \citep{Matignon2007}.
For a temporal-difference error $\delta$, the base learning rate $\alpha$, and $\beta < \alpha$:

\begin{equation}
    \begin{cases}
      \hat{Q}(s,a)\leftarrow \hat{Q}(s,a) + \alpha\delta & \text{if $\delta > 0$}\\
      \hat{Q}(s,a)\leftarrow \hat{Q}(s,a) + \beta\delta & \text{otherwise}\\
    \end{cases}       
\end{equation}

\subsubsection{Supervised loss} 
Inspired by \citep{Hester2018}, we propose a mixed \emph{optimisation-informed centralised but factored critic} approach that incorporates a supervised loss, so agents can learn from both demonstrator and exploration data. 
The supervised loss enables the agent to learn to mimic the expert demonstrations (the convex optimisation results in our problem), while the temporal difference loss enables the agent to learn from its own experience generated through directly interacting with the environment. This mechanism thus combines \say{expensive} demonstration data from convex optimisations and \say{cheap} exploration data from simulation environment to guide the learning of multi-agent cooperation. A weighted penalty is added to the loss for actions that deviate from the demonstrator data:
\begin{equation}
    \delta_\textbf{supervised loss} = C\sum_t{\left(a^t_{i,\textbf{demonstrator}} - a^t_{i,\textbf{exploration}}\right)^2}
\end{equation}
We investigate whether this could guide the agents' learning and reduce the number of exploration steps required, especially as the exploration space and coordination challenges are particularly potent in MARL.

\subsection{RL formulation}
Here, we present how the decision variables in \Cref{sec:system} translate into RL experience tuples $(o_i^t, a_i^t, r^t)$.

\subsubsection{Observation}
The IQL methodology performed best with a small observation space, which we therefore limit to the current grid cost coefficient. In the factored critic methodology, the observation space consists of the grid cost coefficients for 24 hours ahead 



\subsubsection{Actions}
The RL agent selects three different actions. The constraints are then enforced in the environment thanks to a physics-informed approach to translate these agent actions into feasible EV, heating and flexible consumption variables.

\begin{itemize}
\item Battery action $a_{\textrm{bat}, i}^t \in [-1, 1]$: for negative values scales the maximum amount of feasible discharge ($b_{\textrm{out},i}^t$), and for positive values of feasible charge ($b_{\textrm{in},i}^t$), as illustrated in \Cref{fig:battery_action}. Given \Cref{eq:bat_balance,eq:bat_0,eq:bat_bounds,eq:bat_in,eq_bat_out}:
\begin{equation}
\begin{multlined}
    (\underline{\Delta s_i^t}, \overline{\Delta s_i^t}) = \\
    f(E_i^t, d^t_{\textrm{EV}, i}, ..., d^{t_{\textrm{end}}}_{\textrm{EV}, i}, \upmu_i^t, ..., \upmu_i^{t_\textrm{end}}, \textrm{E}_0, \underline{\textrm{E}}, \overline{\textrm{E}}, \overline{\textrm{b}_\textrm{out}}, \overline{\textrm{b}_\textrm{in}})
\end{multlined}
\end{equation}

\begin{figure}[h]
\begin{center}
\includegraphics[width=0.8\linewidth]{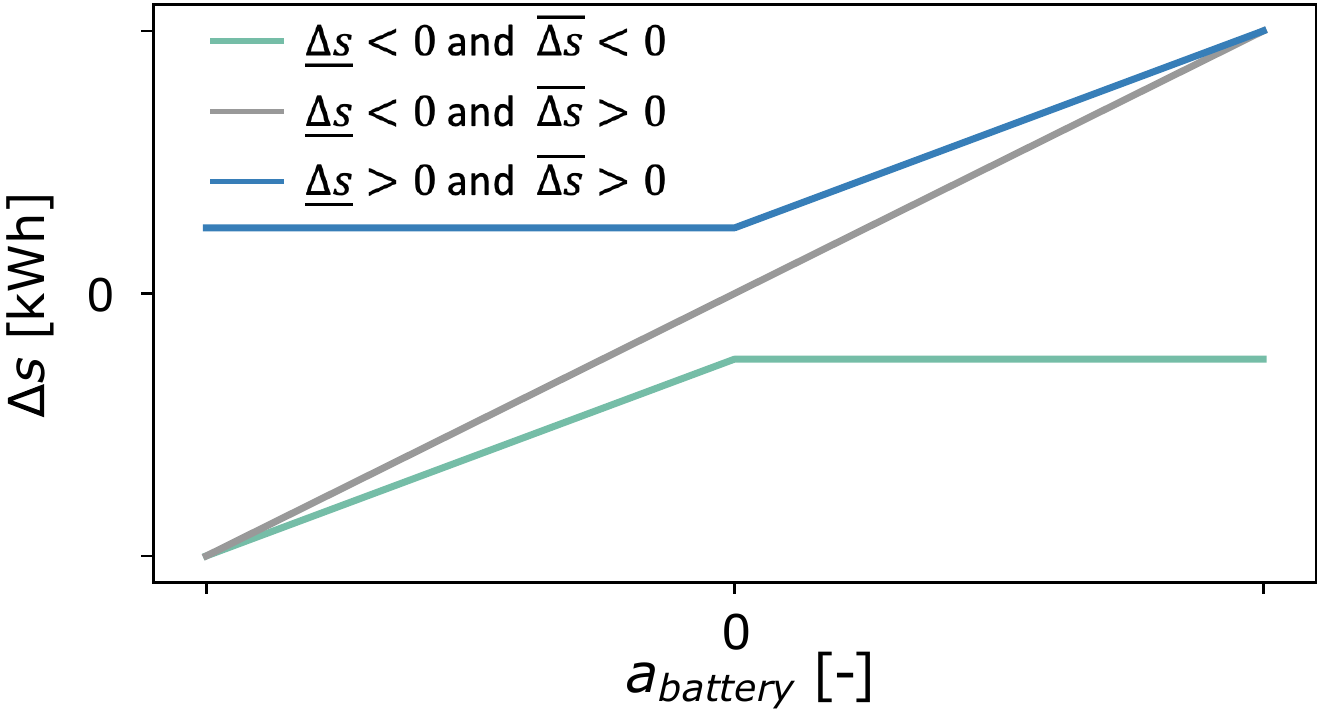}
\end{center}
\caption{Translating the battery action into the change in battery energy level.}
\label{fig:battery_action}
\end{figure}

\item Heating action $a_{\textrm{heat}, i}^t \in [0, 1]$: linearly scales between the minimum and maximum amount of feasible heating ($h_i^t$) given local temperature constraints. Given \Cref{eq:main_heating,eq:temp_limits}:
\begin{equation}
(\underline{h_i^t}, \overline{h_i^t}) = f(T_{\textrm{m},i}^{t}, \textrm{T}_{\textrm{e}}^t, \upphi^t, \overline{T^t_{i,\textrm{air}}}, \underline{T^t_{i,\textrm{air}}})
\end{equation}
\begin{equation}
h_i^t = (1 - a_{i, \textrm{heat}}^t)\underline{h_i^t} +  a_{i, \textrm{heat}}^t\overline{h_i^t}
\end{equation}

\item Household consumption action $a_{\textrm{cons}, i}^t \in [0, 1]$: scales feasible flexible household consumption ($c_i^t$). Given \Cref{eq:totalcons,eq:demandmet}:
\begin{equation}
(\underline{c_i^t}, \overline{c_i^t}) = f(d^{t-\textrm{n}_\textrm{flex}}_i, ..., d^t_i, c_i^{t-\textrm{n}_\textrm{flex}}, ..., c_i^{t-1})
\end{equation}
\begin{equation}
c_i^t = (1 - a_{i, \textrm{cons}}^t)\underline{c_i^t} + a_{i, \textrm{cons}}^t \overline{c_i^t}
\end{equation}
\end{itemize}

\subsubsection{Reward}
The global reward $r^t \in \mathcal{R}$ the RL agents receive at each timestep corresponds to the share $\hat{F}_t$ of the system objective function presented in \Cref{sec:objfunc}.

\section{Results}\label{sec:results}
We conduct a series of experiments to assess the performance and computational scalability of the MARL approaches presented in \Cref{sec:methodology}, when applied to solve the residential flexibility coordination problem in \Cref{sec:system}. Both the optimisation-informed independent learning and the centralised but factored critic approach achieve coordination at scale for the homes controlling their DERs. However, the factored critic achieves this with a nearly 40-fold computational cost reduction for 30 homes.

\subsection{Set-up}

The input data generation is presented in \citep{Charbonnier2022_MARL}. It draws on large-scale real-life data \citep{TC1a, TC5, DepartmentforTransport2019} to generate realistic training and testing samples in a consistent manner. The social cost of carbon is set at 70 £/tCO$_2$, consistent with the UK 2030 target \citep{Hirst2018}. Weather \citep{MERRA2015}, electricity time-of-use prices \citep{OctopusEnergy2019} and grid carbon intensity \citep{NationalGridESO2020} are from January 2021, where relevant specified for Oxford, UK. The low solar heat gains in January are neglected \citep{Brown2020}. EVs are assumed to have a capacity of 39 kWh and 6.6 kW maximum charging rate \citep{nissan}. All code and other parameters are freely available on GitHub\footnote{\url{https://github.com/floracharbo/MARL_local_electricity}}.

\subsection{Results and discussion}
\begin{figure}[!t]
\begin{center}
\includegraphics[width=0.8\linewidth]{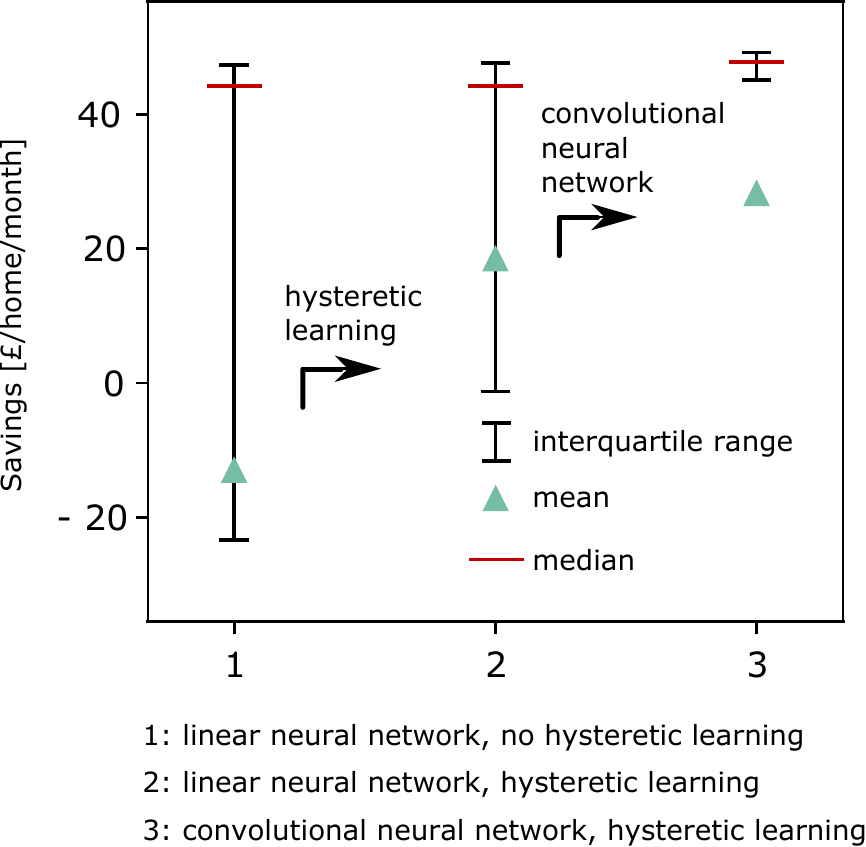}
\end{center}
\caption{Improvements in savings per home and month achieved when adding the use of hysteretic learning and of convolutional neural networks for 10 homes using the adapted FACMAC approach.}
\label{fig:ablations}
\end{figure}

\Cref{fig:ablations} shows that the methodological additions of hysteretic learning and convolutional layers in the actor networks significantly improved the performance of the FACMAC approach when applied to the problem of DER coordination and planning for day-long time series. The increase in average savings is seen to be primarily driven by an improvement of lower percentile \say{worst-case} performances. 

\begin{figure}[h]
\begin{center}
\includegraphics[width=\columnwidth]{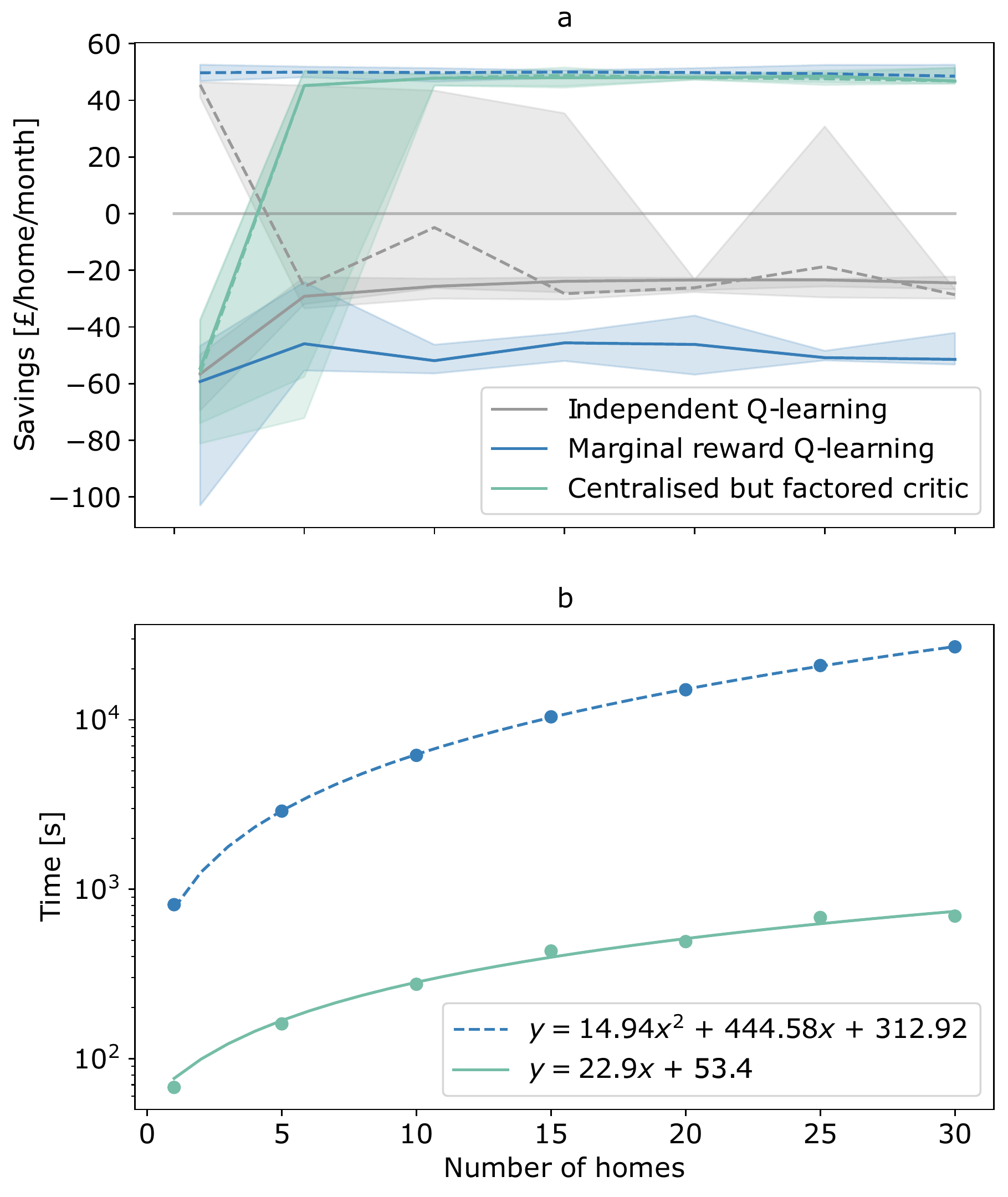}
\end{center}
\caption[(a) 50th percentile savings per home and month relative to the baseline scenario as the number of agents increases over ten repeats. Full lines correspond to training using random environment explorations only. Dotted lines use experience from convex optimisations for training. The shaded areas show the spread between the 25th and 75th percentile values over the ten repeats. (b) Corresponding required computation time for the two methodologies which achieve coordination at scale, namely optimisation-informed independent Q-learning and the centralised but factored critic approach. Functions are fitted using SciPy's optimize.curve\_fit non-linear least squares function \citep{2020SciPy})]{(a) 50th percentile savings per home and month relative to the baseline scenario as the number of agents increases over ten repeats. Full lines correspond to training using random environment explorations only. Dotted lines use experience from convex optimisations for training. The shaded areas show the spread between the 25th and 75th percentile values over the ten repeats. (b) Corresponding required computation time for the 10 repeats for the two methodologies which achieve coordination at scale, namely optimisation-informed independent Q-learning and the centralised but factored critic approach. Functions are fitted using SciPy's optimize.curve\_fit non-linear least squares function \citep{2020SciPy}\protect\footnotemark[6].}
\label{fig:results}
\end{figure}

\begin{figure}[h]
\begin{center}
\includegraphics[width=0.8\linewidth]{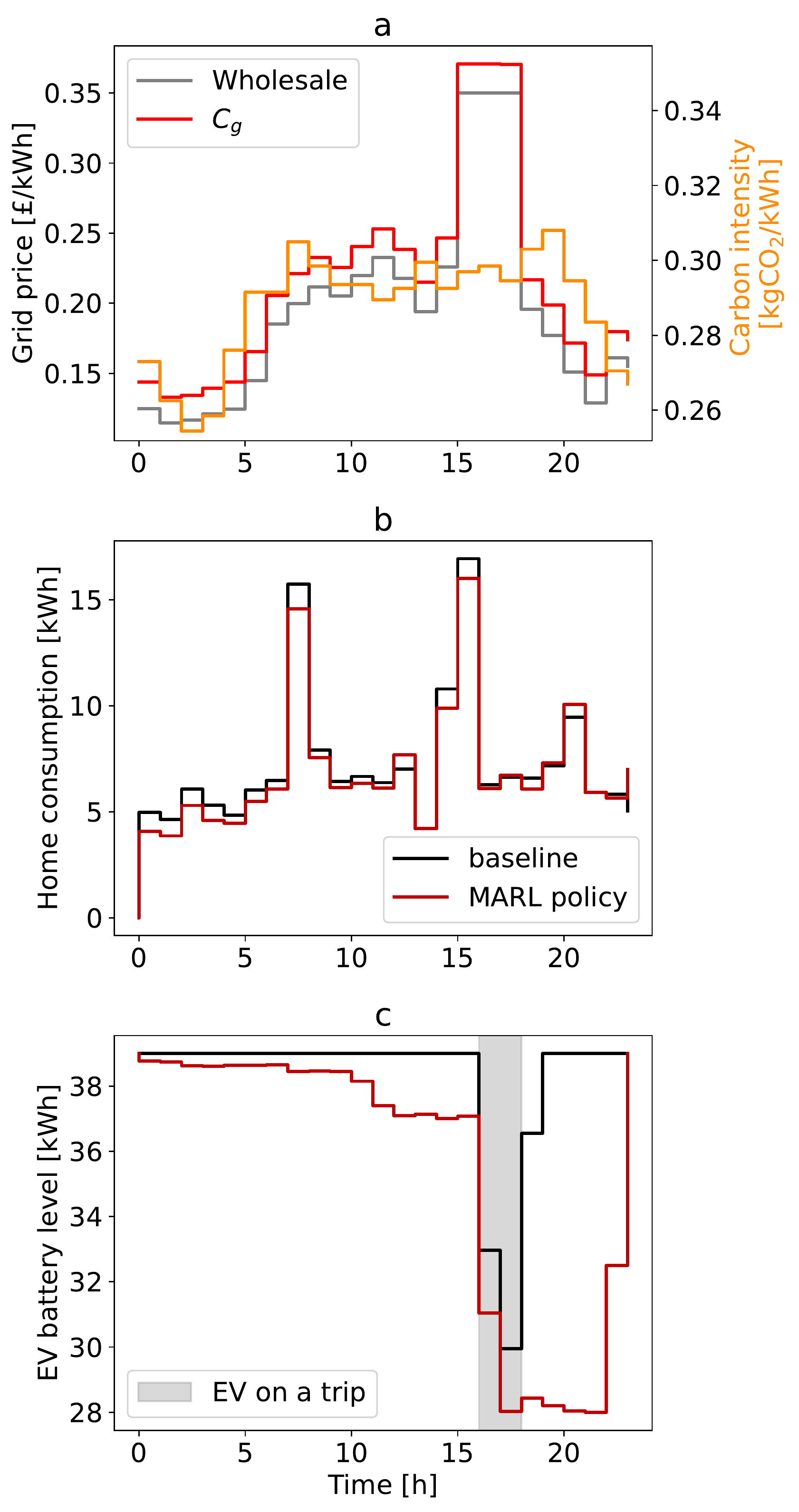}
\end{center}
\caption{Example day: (a) Wholesale prices, grid carbon intensity, and resulting grid cost coefficient $C_g$ given a social cost of carbon of 70 £/tCO$_2$. (b) Total energy consumption, including both household loads and heating consumption. (c) EV battery energy profile.}
\label{fig:example_day}
\end{figure}

Adopting these two additions, \Cref{fig:results}(a) then shows the total system savings achieved per home and month as the number of agents trained increases. These savings are computed relative to a baseline where all agents are inflexible, with EVs charged immediately and no flexible loads delayed. An example day is illustrated in \Cref{fig:example_day}, showing local control variables for a home in the inflexible and coordinated scenarios.

We can see that the coordination performance of IQL (grey line) drops as the number of homes increases, even if learning from optimisation data (grey dotted line). This is due to the learnability issues mentioned above for independent learners. Without a coordination mechanism, independent learners under partial observability in a stochastic environment face challenges such as the incompatibility of individual policy equilibriums with a global Pareto optimal, the non-stationarity of the environment due to other concurrently learning agents affects convergence, and the stochasticity of the environment prevents agents from discriminating between their own contribution to global rewards and noise from other agents or the environment \citep{Matignon2012}.

Only when combining optimisation-informed learning with individual marginal rewards can coordination be achieved at scale (blue dotted line). However, even in cases where optimisation models are available (which is often not the case in complex real-life problems), optimisations are computationally expensive. Indeed, in the interior point method, which is used to solve non-linear continuous problems in convex optimisation solvers, the inverse of the Jacobian of the Karush–Kuhn–Tucker (KKT) conditions must be computed in each Newton-Raphson update step. As this includes derivatives with respect to all decision variables of the problem for all constraints and the objective function \citep{Wright1997}, the Jacobian grows with O($n^2$). This step therefore bears a high computational burden which can be prohibitive as the system size grows \citep{baker2020}, particularly in machine learning applications where numerous optimisations must be performed to generate training data.

The centralised but factored critic methodology (green lines) achieves equivalent coordination performance to the optimisation-informed marginal reward Q-learning approach, whilst overcoming its dependency on optimisations and additional marginal rewards computations. The objective function in \Cref{sec:system} could be reduced by £49.84 per home and month on average for 30 homes, or a 7.4\% reduction. £46.51 was saved through reduced consumer energy bills and £3.56 through greenhouse gas emissions reduction. At the same time, battery depreciation costs increased by £0.17 and distribution network use increased the total costs by £0.05 per month and home relative to the inflexible baseline.

The centralised but factored critic can capture a global understanding of the impact of individual policies on the system via the centralised action-value function, using extra information (e.g., global state and joint action) available only during training. In the residential DER coordination problem, there is particularly high value in cooperation signals that take a global view of the system, as the global rewards are highly dependent on the cumulative impact of multiple agents taking actions simultaneously. The backward propagation of gradients from the centralised but factored critic to the individual actor networks can then guide the optimisation of individual policies in the absence of full state information at the scale of individual agents \citep{Papoudakis2020}. FACMAC can also enable more scalable learning as the centralised action-value function is represented as a combination of individual action-value functions, which condition on much smaller local observations and actions. The factored critic uses information more efficiently as network weight updates use state space information that has the most impact on the global value of actions taken by agents. The updates structurally take into account the partial observability of agents taking actions, and aim to update only the knowledge necessary to know which actions agents should take and when. 

\footnotetext[6]{Fitting a second-order polynomial function into the computational time of the centralised but factored critic approach yields a second-order term coefficient of -0.175 and a first-order coefficient of 28.272. This confirms the hypothesis that a first-order complexity representation is a better fit, as illustrated in \Cref{fig:results}(b).}

The back-propagation from the centralised but factored critic thus mirrors the global optimisation, given the partial observability of the agents taking actions: each back-propagation includes a Jacobian-gradient product of the value error with respect to the weights of the networks for each operation in the graph. \Cref{fig:results} thus shows that optimisation-informed FACMAC (incorporating the supervised loss, green dotted line) does not achieve superior performance relative to FACMAC. This demonstrates that the centralised but factored critic alone already provides an equivalent coordination mechanism for the cooperation of the agents in the system, and the use of optimisation data adds no additional value. 

The optimisation-informed independent learning and the centralised but factored critic approaches thus mirror each other in their provision of a global coordination mechanism and achieve very similar performance, while having structural differences which lead to varying computational efficiencies. In both cases, the Jacobian plays a role, and both marginal rewards and the backward propagation from the centralised but factored critic aim to improve learnability, sending personalised rewards to each agent that best represent their contribution to the global reward. One advantage of IQL is that it does not require neural networks and, as such, is more robust to hyper-parameter tuning and more easily interpretable, provided it can learn from model optimisation results. It also provides superior coordination performance below 10 agents, as FACMAC was designed for scale, when more invidual critic networks can yield better centralised critic factorisation. Nevertheless, the factored critic uses information more efficiently and achieves superior computational scalability as the number of value networks only grows linearly with the number of homes O($n$). \Cref{fig:results}(b) shows the optimisation-informed approach required second-order polynomial computational time as the number of agents increases, whereas the centralised but factored critic methodology required approximately first-order polynomial time. For 30 homes, the factored critic approach thus required 38.9 times less computational time.

\section{Conclusion}\label{sec:conclusion}
This paper has proposed the use of a cooperative multi-agent reinforcement learning approach that learns a centralised but factored critic to enable privacy-preserving and scalable coordination of residential energy flexibility.

Compared with a previous state-of-the-art independent learning approach which learned from global optimisation results, the centralised but factored critic  yielded similar coordination at scale, as each coordinated home provided an average reduction of £46.82 in global system costs per month for 30 homes. Moreover, the centralised but factored critic approach required computational time nearly 40 times lower for 30 homes, and growing only at first-order rather than second-order polynomial time. This improved scalability opens the way for coordinating flexible energy resources such as electric vehicles and heating in a fully distributed manner.

The impact of such coordination on the distribution network and the potential cooperative management of network constraints by agents could be further investigated.




\begin{thebibliography}{10}

\bibitem{Torriti2022}
J.~Torriti.
\newblock {Household electricity demand, the intrinsic flexibility index and UK
  wholesale electricity market prices}.
\newblock {\em Environmental Economics and Policy Studies}, 24(1):7--27, 2022.

\bibitem{Leautier2019}
T-O. L{\'{e}}autier.
\newblock {\em {Imperfect Markets and Imperfect Regulation: An Introduction to
  the Microeconomics and Political Economy of Power Markets}}.
\newblock MIT Press, 2019.

\bibitem{Pumphrey2020}
K.~Pumphrey, S.~Walker, M.~Andoni, and V.~Robu.
\newblock {Green hope or red herring? Examining consumer perceptions of
  peer-to-peer energy trading in the United Kingdom}.
\newblock {\em Energy Research and Social Science}, 68(September 2019):101603,
  2020.

\bibitem{Bugden2019}
D.~Bugden and R.~Stedman.
\newblock {A synthetic view of acceptance and engagement with smart meters in
  the United States}.
\newblock {\em Energy Research and Social Science}, 47(January 2018):137--145,
  2019.

\bibitem{Charbonnier2022_Review}
F.~Charbonnier, T.~Morstyn, and M.~Mcculloch.
\newblock {Coordination of resources at the edge of the electricity grid :
  Systematic review and taxonomy}.
\newblock {\em Applied Energy}, 318(April):119188, 2022.

\bibitem{ONeill2010}
D.~O'Neill, M.~Levorato, A.~Goldsmith, and U.~Mitra.
\newblock {Residential Demand Response Using Reinforcement Learning}.
\newblock {\em 2010 First IEEE International Conference on Smart Grid
  Communications}, pages 409--414, 2010.

\bibitem{Antonopoulos2020}
{I. Antonopoulos et al.}
\newblock {Artificial intelligence and machine learning approaches to energy
  demand-side response: A systematic review}.
\newblock {\em Renewable and Sustainable Energy Reviews}, 130(April):109899,
  2020.

\bibitem{Darby2020}
S.~Darby.
\newblock {Demand response and smart technology in theory and practice:
  Customer experiences and system actors}.
\newblock {\em Energy Policy}, 143(April):111573, 2020.

\bibitem{Chen2019}
T.~Chen and W.~Su.
\newblock {Indirect Customer-to-Customer Energy Trading with Reinforcement
  Learning}.
\newblock {\em IEEE Transactions on Smart Grid}, 10(4):4338--4348, 2019.

\bibitem{Lu2019}
R.~Lu and S.~Hong.
\newblock {Incentive-based demand response for smart grid with reinforcement
  learning and deep neural network}.
\newblock {\em Applied Energy}, 236(December 2018):937--949, 2019.

\bibitem{Kim2016}
B.~Kim, Y.~Zhang, M.~Van Der~Schaar, and J.~Lee.
\newblock {Dynamic Pricing and Energy Consumption Scheduling With Reinforcement
  Learning}.
\newblock {\em IEEE Transactions on Smart Grid}, 7(5):2187--2198, 2016.

\bibitem{Babar2018}
P.~H. Babar, M. 
  Zbigniew Hanzelka.
\newblock {The evaluation of agile demand response: An applied methodology}.
\newblock {\em IEEE Transactions on Smart Grid}, 9(6):6118--6127, 2018.

\bibitem{Vaya2014}
M.~G. Vay{\'{a}}, L.~B. Rosell{\'{o}}, and G.~Andersson.
\newblock {Optimal bidding of plug-in electric vehicles in a market-based
  control setup}.
\newblock {\em Proceedings - 2014 Power Systems Computation Conference, PSCC
  2014}, 2014.

\bibitem{Ye2020}
Y.~Ye, D.~Qiu, M.~Sun, D.~Papadaskalopoulos, and G.~Strbac.
\newblock {Deep Reinforcement Learning for Strategic Bidding in Electricity
  Markets}.
\newblock {\em IEEE Transactions on Smart Grid}, 11(2):1343--1355, 2020.

\bibitem{Dauer2013}
D.~Dauer, C.~Flath, P.~Str{\"{o}}hle, and C.~Weinhardt.
\newblock {Market-based EV charging coordination}.
\newblock {\em Proceedings - 2013 IEEE/WIC/ACM International Conference on
  Intelligent Agent Technology, IAT 2013}, 2:102--107, 2013.

\bibitem{Sun2015}
Y.~Sun, A.~Somani, and T.~Carroll.
\newblock {Learning based bidding strategy for HVAC systems in double auction
  retail energy markets}.
\newblock {\em Proceedings of the American Control Conference},
  2015-July:2912--2917, 2015.

\bibitem{Kim2020}
J.~Kim and B.~Lee.
\newblock {Automatic P2P energy trading model based on reinforcement learning
  using long short-term delayed reward}.
\newblock {\em Energies}, 13(20), 2020.

\bibitem{Claessens2013}
B.~Claessens, S.~Vandael, F.~Ruelens, K.~{De Craemer}, and B.~Beusen.
\newblock {Peak shaving of a heterogeneous cluster of residential flexibility
  carriers using reinforcement learning}.
\newblock {\em 2013 4th IEEE/PES Innovative Smart Grid Technologies Europe,
  ISGT Europe 2013}, pages 1--5, 2013.

\bibitem{Zhang2017}
X.~Zhang, T.~Bao, T.~Yu, B.~Yang, and C.~Han.
\newblock {Deep transfer Q-learning with virtual leader-follower for
  supply-demand Stackelberg game of smart grid}.
\newblock {\em Energy}, 133:348--365, 2017.

\bibitem{Dusparic2015}
I.~Dusparic, A.~Taylor, A.~Marinescu, V.~Cahill, and S.~Clarke.
\newblock {Maximizing renewable energy use with decentralized residential
  demand response}.
\newblock {\em 2015 IEEE 1st International Smart Cities Conference, ISC2 2015},
  2015.

\bibitem{Hurtado2018}
L.~Hurtado, E.~Mocanu, P.~Nguyen, M.~Gibescu, and R.~Kamphuis.
\newblock {Enabling Cooperative Behavior for Building Demand Response Based on
  Extended Joint Action Learning}.
\newblock {\em IEEE Transactions on Industrial Informatics}, 14(1):127--136,
  2018.

\bibitem{Taylor2014}
A.~Taylor, I.~Dusparic, E.~Galvan-Lopez, S.~Clarke, and V.~Cahill.
\newblock {Accelerating Learning in multi-objective systems through Transfer
  Learning}.
\newblock {\em Proceedings of the International Joint Conference on Neural
  Networks}, pages 2298--2305, 2014.

\bibitem{Guerrero2020}
J.~Guerrero, D.~Gebbran, S.~Mhanna, A.~Chapman, and G.~Verbič.
\newblock Towards a transactive energy system for integration of distributed
  energy resources: Home energy management, distributed optimal power flow, and
  peer-to-peer energy trading.
\newblock {\em Renewable \& sustainable energy reviews}, 132, 2020.

\bibitem{Morstyn2020_P2P}
T.~Morstyn and M.~Mcculloch.
\newblock {Peer-to-Peer Energy Trading}.
\newblock {\em Analytics for the Sharing Economy: Mathematics, Engineering and
  Business Perspectives}, (March), 2020.

\bibitem{Mai2021}
{T. Mai et al.}
\newblock {An overview of grid-edge control with the digital transformation}.
\newblock {\em Electrical Engineering}, 103(4):1989--2007, 2021.

\bibitem{oliehoek2016concise}
Frans~A Oliehoek and Christopher Amato.
\newblock {\em A concise introduction to decentralized POMDPs}.
\newblock Springer, 2016.

\bibitem{Cao2019}
J.~Cao.
\newblock {Deep Reinforcement Learning Based Energy Storage Arbitrage With
  Accurate Lithium-ion Battery Degradation Model}.
\newblock {\em IEEE Transactions on Smart Grid}, 14(8):1--9, 2019.

\bibitem{Yang2019}
Y.~Yang, J.~Hao, Y.~Zheng, and C.~Yu.

\bibitem{Crozier2018_Mitigating}
C.~Crozier, D.~Apostolopoulou, and M.~McCulloch.
\newblock {Mitigating the impact of personal vehicle electrification: A power
  generation perspective}.
\newblock {\em Energy Policy}, 118(2013):474--481, 2018.

\bibitem{Pigott2022}
A.~Pigott, C.~Crozier, K.~Baker, and Z.~Nagy.
\newblock {GridLearn: Multiagent reinforcement learning for grid-aware building
  energy management}.
\newblock {\em Electric Power Systems Research}, 213(October 2021):108521,
  2022.

\bibitem{Charbonnier2022_MARL}
F.~Charbonnier, T.~Morstyn, and M.~McCulloch.
\newblock Scalable multi-agent reinforcement learning for distributed control
  of residential energy flexibility.
\newblock {\em Applied Energy}, 314:118825, 2022.

\bibitem{oliehoek2008optimal}
F.~A Oliehoek, M.~Spaan, and N.~Vlassis.
\newblock Optimal and approximate q-value functions for decentralized pomdps.
\newblock {\em Journal of Artificial Intelligence Research}, 32:289--353, 2008.

\bibitem{Tan1993}
M.~Tan.
\newblock {Multi-Agent Reinforcement Learning : Independent vs. Cooperative
  Agents}.
\newblock 1993.

\bibitem{Rashid2018}
T.~Rashid, M.~Samvelyan, C.~Schroeder de~Witt, G.~Farquhar, J.~Foerster, and
  S.~Whiteson.
\newblock {QMIX: Monotonic Value Function Factorisation for Deep Multi-Agent
  Reinforcement Learning Tabish}.
\newblock {\em Proceedings of the 35th International Conference on Machine
  Learning}, 2018.

\bibitem{Matignon2012}
L.~Matignon, G.~Laurent, and N.~{Le Fort-Piat}.
\newblock {Independent reinforcement learners in cooperative Markov games: A
  survey regarding coordination problems}.
\newblock {\em Knowledge Engineering Review}, 27(1):1--31, 2012.

\bibitem{Gupta2017}
J.~Gupta, M.~Egorov, and M.~Kochenderfer.
\newblock {Cooperative Multi-agent Control Using Deep Reinforcement Learning}.
\newblock {\em Lecture Notes in Computer Science (including subseries Lecture
  Notes in Artificial Intelligence and Lecture Notes in Bioinformatics)}, 10642
  LNAI:66--83, 2017.

\bibitem{Foerster2018}
J.~Foerster, G.~Farquhar, T.~Afouras, N.~Nardelli, and S.~Whiteson.
\newblock {Counterfactual multi-agent policy gradients}.
\newblock {\em 32nd AAAI Conference on Artificial Intelligence, AAAI 2018},
  pages 2974--2982, 2018.

\bibitem{Lowe2017}
R.~Lowe, Y.~Wu, A.~Tamar, J.~Harb, A.~Pieter, and I.~Mordatch.
\newblock Multi-agent actor-critic for mixed cooperative-competitive
  environments.
\newblock In I.~Guyon, U.~Von Luxburg, S.~Bengio, H.~Wallach, R.~Fergus,
  S.~Vishwanathan, and R.~Garnett, editors, {\em Advances in Neural Information
  Processing Systems}, volume~30. Curran Associates, Inc., 2017.

\bibitem{Lowe2020}
R.~Lowe.
\newblock {Multi-Agent Actor-Critic for Mixed Cooperative-Competitive
  Environments}.
\newblock 2020.

\bibitem{rashid2020weighted}
T.~Rashid, G.~Farquhar, B.~Peng, and S.~Whiteson.
\newblock Weighted {QMIX}: Expanding monotonic value function factorisation for
  deep multi-agent reinforcement learning.
\newblock In {\em Advances in Neural Information Processing Systems},
  volume~33, pages 10199--10210, 2020.

\bibitem{wang2021qplex}
J.~Wang, Z.~Ren, T.~Liu, Y.~Yu, and C.~Zhang.
\newblock {QPLEX}: Duplex dueling multi-agent {Q}-learning.
\newblock In {\em International Conference on Learning Representations}, 2021.

\bibitem{Hu2020}
Jian H., Seth~A. H., Haibin W., and Shih{-}wei L.
\newblock {QR-MIX:} distributional value function factorisation for cooperative
  multi-agent reinforcement learning.
\newblock {\em CoRR}, abs/2009.04197, 2020.

\bibitem{Qiu2021}
{W. Qiu, et al.}
\newblock Rmix: Learning risk-sensitive policies for cooperative reinforcement
  learning agents.
\newblock 2021.

\bibitem{Papoudakis2020}
G.~Papoudakis, F.~Christianos, L.~Schäfer, and S.~V. Albrecht.
\newblock Benchmarking multi-agent deep reinforcement learning algorithms in
  cooperative tasks, 2020.

\bibitem{Mnih2013}
{V. Mnih et al.}
\newblock Playing atari with deep reinforcement learning.
\newblock {\em CoRR}, abs/1312.5602, 2013.

\bibitem{Omidshafiei2017}
S.~Omidshafiei, J.~Pazis, C.~Amato, J.~P. How, and J.~Vian.
\newblock {Deep decentralized multi-task multi-agent reinforcement learning
  under partial observability}.
\newblock {\em 34th International Conference on Machine Learning, ICML 2017},
  6:4108--4122, 2017.

\bibitem{Foerster2017}
{J. Foerster et al.}
\newblock {Stabilising experience replay for deep multi-agent reinforcement
  learning}.
\newblock {\em 34th International Conference on Machine Learning, ICML 2017},
  3:1879--1888, 2017.

\bibitem{Peng2021}
{B. Peng et al.}
\newblock Facmac: Factored multi-agent centralised policy gradients.
\newblock 34:12208--12221, 2021.

\bibitem{Lecun1998}
Y.~Lecun, L.~Bottou, Y.~Bengio, and P.~Haffner.
\newblock Gradient-based learning applied to document recognition.
\newblock {\em Proceedings of the IEEE}, 86(11):2278--2324, 1998.

\bibitem{Matignon2007}
L.~Matignon, G.~Laurent, and N.~{Le Fort-piat}.
\newblock {Hysteretic Q-Learning : an algorithm for Decentralized Reinforcement
  Learning in Cooperative Multi-Agent Teams.}
\newblock In {\em Proceedings of the 2007 IEEE/RSJ International Conference on
  Intelligent Robots and Systems}, pages 64--69. IEEE, 2007.

\bibitem{Hester2018}
{T. Hester et al.}
\newblock {Deep q-learning from demonstrations}.
\newblock {\em 32nd AAAI Conference on Artificial Intelligence, AAAI 2018},
  pages 3223--3230, 2018.

\bibitem{ParryM}
M.~Parry.
\newblock {\em Climate change 2007: impacts, adaptation and vulnerability}.
\newblock Published for the Intergovernmental Panel on Climate Change [by]
  Cambridge University Press, Cambridge, 2007.

\bibitem{Morstyn2020_IntegratingP2P}
T.~Morstyn, A.~Teytelboym, C.~Hepburn, and M.~McCulloch.
\newblock {Integrating P2P Energy Trading with Probabilistic Distribution
  Locational Marginal Pricing}.
\newblock {\em IEEE Transactions on Smart Grid}, 11(4):3095--3106, 2020.

\bibitem{DufoLopez2014}
R.~Dufo-L{\'{o}}pez, J.~M Lujano-Rojas, and J.~Bernal-Agustín.
\newblock Comparison of different lead–acid battery lifetime prediction
  models for use in simulation of stand-alone photovoltaic systems.
\newblock {\em Applied energy}, 115:242--253, 2014.

\bibitem{ISO2007}
ISO.
\newblock {Calculation of Energy Use for Space Heating and Cooling ISO/FDIS
  13790:2007(E)}, 2007.

\bibitem{Sutton1998}
R.~Sutton and A.~Barto.
\newblock {\em Reinforcement learning : an introduction [electronic resource]}.
\newblock Adaptive computation and machine learning. MIT Press, Cambridge,
  Mass., 1998.

\bibitem{TC1a}
R.~Wardle.
\newblock {Dataset (TC1a): Basic Profiling of Domestic Smart Meter Customers},
  2014.

\bibitem{TC5}
R.~Wardle.
\newblock {Dataset (TC5): Enhanced Profiling of Domestic Customers with Solar
  Photovoltaics (PV)}, 2014.

\bibitem{DepartmentforTransport2019}
{Department for Transport}.
\newblock {National Travel Survey 2002-2017}, 2019.

\bibitem{Hirst2018}
D.~Hirst.
\newblock {Commons Briefing Paper SNO5927: Carbon Price Floor (CPF) and the
  price support mechanism}, 2018.

\bibitem{MERRA2015}
Global Modeling and Assimilation~Office (GMAO).
\newblock Merra-2 {inst1\_2d\_asm\_Nx}: 2d,1-hourly, instantaneous,
  single-level, assimilation, single-level diagnostics v5.12.4, 2015.

\bibitem{OctopusEnergy2019}
{Octopus Energy}.
\newblock {Octopus Energy API}, 2019.

\bibitem{NationalGridESO2020}
{National Grid ESO}, {Environmental Defense Fund Europe}, {University of Oxford
  Department of Computer Science}, and WWF.
\newblock {Carbon Intensity API}, 2020.

\bibitem{Brown2020}
J.~Brown, J.~Chambers, and A.~Rogers.
\newblock {SMITE : Using Smart Meters to Infer the Thermal Efficiency of
  Residential Homes}.
\newblock In {\em The 7th ACM International Conference on Systems for Energy-
  Efficient Buildings, Cities, and Transportation (BuildSys '20)}, 2020.

\bibitem{nissan}
Nissan~Intelligent Mobility.
\newblock Nissan leaf, 2022.

\bibitem{2020SciPy}
P.~Virtanen.
\newblock {SciPy 1.0: Fundamental Algorithms for Scientific Computing in
  Python}.
\newblock {\em Nature Methods}, 17:261--272, 2020.

\bibitem{Wright1997}
S.~Wright.
\newblock {\em Primal-Dual Interior-Point Methods}.
\newblock Society for Industrial and Applied Mathematics, 1997.

\bibitem{baker2020}
K.~Baker.
\newblock A learning-boosted quasi-newton method for ac optimal power flow.
\newblock {\em arXiv preprint arXiv:2007.06074}, 2020.

\end{thebibliography}

\newpage


\vfill

\end{document}